\documentclass[twocolumn]{autart}

\usepackage{graphicx}
\usepackage{subcaption}
\graphicspath{{Figures/}}

\usepackage{amsmath}  
\usepackage{amssymb}
\usepackage{bm}

\usepackage{color}
\usepackage{cite}
\usepackage{array}

\newcommand{\eqn}[1]{Eqn.~(\ref{#1})}
\newcommand{\eqns}[1]{Eqns.~(\ref{#1})}
\newcommand{\eqnref}[1]{(\ref{#1})}
\newcommand{\fig}[1]{Fig.~\ref{#1}}
\newcommand{\figs}[1]{Figs.~\ref{#1}}
\newcommand{\tab}[1]{Table~\ref{#1}}

\newtheorem{theorem}{Theorem}
\newtheorem{lemma}{Lemma}
\newtheorem{remark}{Remark}
\newtheorem{proposition}{Proposition}
\newtheorem{definition}{Definition}
\newtheorem{corollary}{Corollary}

\begin{document}
\begin{frontmatter}
\title{A New Exponential Forgetting Algorithm for Recursive Least-Squares Parameter Estimation}

\thanks[footnoteinfo]{This paper was not presented at any IFAC 
meeting. Corresponding author Hyo-Sang Shin. Tel. +44-1234-758577. }

\author[Cran]{Hyo-Sang Shin}\ead{h.shin@cranfield.ac.uk},
\author[Cran]{Hae-In Lee}\ead{haein.lee@cranfield.ac.uk} 

\address[Cran]{Cranfield University, College Rd, Cranfield, Bedford MK43 0AL, UK}
          

\begin{abstract}
This paper develops a new exponential forgetting algorithm that can prevent so-called the estimator windup problem, while retaining fast convergence speed. To investigate the properties of the proposed forgetting algorithm, boundedness of the covariance matrix is first analysed and compared with various exponential and directional forgetting algorithms. Then, stability of the estimation error with and without the persistent excitation condition is theoretically analysed in comparison with the existing benchmark algorithms. Numerical simulations on wing rock motion validate the analysis results. 
\end{abstract}

\end{frontmatter}

\section{Introduction}
The Recursive Least Square (RLS) approach has been widely implemented in parameter estimation. The aim of the RLS algorithm is to find a recursive algorithm that minimises the sum of squares of the estimation errors, that is, the difference between the observed and estimated values. The issue with the RLS algorithm is that the weight, i.e. the adaptation gain, to the new estimation errors  becomes less: such adaptation gains are known to be inappropriate for estimating time-varying parameters. 

The adaptation gain issue in the RLS algorithm can be resolved, in a certain degree, by discounting the obsolete information. Exponential forgetting (EF) algorithms incorporate a forgetting factor to discount the obsolete information, in favour of new information that is conveyed by recent data \cite{Astrom1977}. The forgetting rate of the obsolete information in such an algorithm is exponential and this is the reason why they are called EF. The forgetting factor is typically constant, but some EF algorithms utilise a variable forgetting factor to better handle time-varying parameters \cite{Zhang2001,Song2000,Hou2018}. 

It is proven that the EF algorithm well behaves under the persistent excitation (PE) condition \cite{Vau2018}. Here, persistent excitation implies that the observed data sequence contains sufficient information in all the parameter space. If the persistent excitation condition does not hold, it is difficult to guarantee stability of the EF algorithm. It is also known that the EF algorithm could suffer from estimator windup without the persistent excitation condition satisfied \cite{Johnstone1982,Astrom1995,Medvedev2004}. In the EF algorithm, the obsolete information is uniformly discounted along with all directions in the parameter space, but only the part excited can be replaced by the incoming data. This could result in unbounded adaptation gains, which make the estimation algorithm sensitive to noise. This phenomenon is known as estimator windup and undesirable in parameter estimation.

If the incoming information is non-uniformly distributed in the parameter space, it would be desirable to perform selective forgetting: forget the obsolete information only when the incoming data can replace them. This concept has been achieved in directional forgetting (DF) \cite{Kulhavy1984}. There have been extensive studies on developing DF algorithms and investigating their performance \cite{Bertin1987,Campi1992,Hagglund1985,Kulhavy1987} and applications on controls \cite{Li2016,Santin2017,Rojas2018}. It is proven that although most of the DF algorithms can prevent estimation windup, some of the eigenvalues of the covariance matrix might become zero in some DF algorithms \cite{Bittanti1990convergence,Bittanti1994}. This means that some components in the adaptation gain become null and thus the tracking capability of those algorithms might be lost in some directions. Cao and Schwartz \cite{Cao1999,Cao2000} developed a DF algorithm based on a matrix decomposition to guarantee the boundedness of the covariance matrix, which prevents the adaptation gain becoming zero. Nonetheless, the stability characteristics of their DF algorithm was not analysed.

A potential issue with the DF algorithm is that its convergence speed is generally slower than that of the EF algorithm. This is because the adaptation gain along with the direction of the obsolete information, which is not replaced by incoming information, becomes relatively small and thus insensitive to the estimation errors. The convergence speed is important, as it indicates adaptation capability to time-varying parameters \cite{Bittanti1990convergence}. This importance becomes paramount in some applications, for example in indirect adaptive control for a system in which fast parameter estimation becomes essential for the control  \cite{Chikasha2017,Cho2017,Salahshoor2015}.

To this end, this paper aims to propose a new EF algorithm that can alleviate the issues with estimator windup and the PE requirement for the stability guarantee, while retaining adaptation capability. The proposed algorithm thus enables relaxation of the limitations of EF and DF algorithms at the same time. To develop such an algorithm, this paper is first devoted to investigate pre-existing forgetting algorithms. In this step, we select representative EF and DF algorithms as a benchmark example for the analysis and discuss potential issues with those algorithms. Then, this paper proposes a new EF algorithm and performs theoretical analysis on its properties. 

The theoretical analysis is conducted on the two main points: the boundedness on the covariance matrix and the stability of the estimation error. It is worth noting that most of theoretical analyses are conducted under the PE condition: the analyses become inconclusive without the PE condition satisfied. Therefore, one main focus of our analysis is to relax the PE condition and perform the analysis. 

The analysis results show that the covariance matrix in the proposed EF algorithm is bounded from above and below even without the PE condition met, whereas the boundedness of the previous EF or some DF algorithms is not guaranteed. Note that the windup issue and adaptation capability have been investigated mainly by checking the boundedness of the covariance matrix. If the covariance matrix is bounded from below, the forgetting algorithm provides a certain level of adaptation capability. If it is bounded from above, the estimator windup issue can be alleviated. Hence, the theoretical analysis results imply that the proposed method can mitigate the windup issue while retaining the adaptation capability. 

The stability characteristics of the proposed EF and other benchmark RLS algorithms are analysed with and also without the PE condition. To the best of our knowledge, most, if not all, of the stability analyses of the RLS-based algorithms are performed under the PE assumption. The analysis results reveal that the proposed EF algorithm guarantees exponential stability with the PE condition satisfied, whereas the DF algorithms guarantee only stability or uniform stability. This paper carries out the stability analysis of the DF algorithm developed in \cite{Cao1999,Cao2000} for the first time for the comparison purpose. It is also proven that the proposed EF algorithm developed can guarantee uniform stability even without the PE condition met, unlike the conventional EF algorithm. Note that like in other RLS algorithms \cite{Bittanti1990convergence,Johnstone1982}, the main analysis of the stability characteristics is performed in a deterministic environment. 

Numerical simulations are conducted with respect to wing rock motion of an aircraft. The numerical results with the proposed EF algorithm are compared with those of the previous EF and DF algorithm, and validate the conformance with the analysis results.

The rest of the paper is organised as follows. Section \ref{sec:preliminaries} provides key preliminaries and background that are essential for the development and analysis of the proposed algorithm. Section \ref{sec:existing} conducts analysis on the existing benchmark EF and DF algorithms. The new EF algorithm is introduced and its properties are theoretically analysed in comparison with those of the benchmark algorithms in Section \ref{sec:newEF}. The properties of the EF algorithm newly developed are demonstrated and compared with the benchmark algorithms in Section \ref{sec:NS}. Section \ref{sec:conclusion} offers conclusions.

\section{Preliminaries}
\label{sec:preliminaries}

The RLS algorithm and its variants are typically expressed as:
\begin{equation}
\hat{\theta} (t)= \hat{\theta} (t-1) + K(t) \phi (t) [ y(t) - \phi^T(t) \hat{\theta} (t-1)], \label{eq1.1}
\end{equation}
where $\hat{\theta}(t)$ is the estimate of the parameter vector at time step $t$, $\phi(t)$ the observed data vector, $y(t)$ the system output vector, and $K(t)$ the adaptation gain. Given a constant unknown parameter vector $\theta$, the output satisfies $y(t)=\phi^T(t)\theta$, and thus \eqn{eq1.1} can be rewritten as:
\begin{equation}
\label{eq1.3}
\tilde{\theta}(t)= \left[ I - K(t) \phi(t) \phi^T(t) \right] \tilde{\theta}(t-1), \\
\end{equation}
where $\tilde{\theta}(t) = \hat{\theta}(t) - \theta$.

The adaptation gain $K(t)$ is time-varying to reflect the error covariance on the adaptation speed, and hence generally a function of the covariance matrix $P(t)$. The dynamics of $P(t)$ is typically given by:
\begin{equation}
R(t) = F(t) R(t-1) + \phi(t) \phi^T (t), \label{eq1.2}
\end{equation}
where $R(t)=P^{-1}(t)$ is the information matrix, and $F(t)$ is the forgetting matrix. It is assumed that the initial covariance and information matrices are or set to be positive definite, i.e. $P(0)>0$ and $R(0)>0$.  

Design of the adaptation gain $K(t)$ and the forgetting matrix $F(t)$ differ in various types of the EF and DF algorithms, depending on how to discount obsolete information. As discussed in Introduction, we selected a few well known forgetting algorithms as a benchmark: EF \cite{Astrom1977}, DF$^1$ \cite{Kulhavy1984}, and DF$^2$ \cite{Cao2000}. The adaptation gain matrices of the selected algorithms can be summarised as:
\begin{equation}
\left\{\begin{aligned}
&\text{EF}: &K(t)&=\cfrac{P(t-1)}{\mu+\phi^T(t) P(t-1)\phi(t)}\\
&\text{DF}^1: &K(t)&=\cfrac{P(t-1)}{1+\phi^T(t) P(t-1)\phi(t)}\\
&\text{DF}^2: &K(t)&=P(t),
\end{aligned}\right.
\end{equation}
and the forgetting matrices as:
\begin{equation}
\left\{\begin{aligned}
&\text{EF}: &F(t)&=\mu I\\
&\text{DF}^1: &F(t)&=I-(1-\beta(t))\phi(t)\phi^T(t) P(t-1)\\
&\text{DF}^2: &F(t)&=I-(1-\mu)\cfrac{P^{-1}(t-1)\phi(t)\phi^T(t)}{\phi^T(t)P^{-1}(t-1)\phi(t)},
\end{aligned}\right.
\label{eq1.4}
\end{equation}
where $\mu\in(0,1)$ is the forgetting factor and $\beta(t)$ is a scalar defined as:
\begin{equation}
\label{eq1.4.5}
\beta(t)=\mu-\frac{1-\mu}{\phi^T(t) P(t-1)\phi(t)}.
\end{equation}

Depending on the value of $\mu$, $\beta(t)$ could become negative. 

For the completeness, we provide the definition of PE, following those from \cite{Tao2003,Bittanti1990convergence}. 

\begin{definition}[PE]
The observed data vector $\phi(t)$ is persistently exciting of order $s$ if there exist $s>0$ and $\gamma >0$ such that
    \begin{equation}
    \sum_{i=0}^{s-1}\phi(t-i)\phi^T(t-i)\geq \gamma I, \qquad \forall t>s.
    \end{equation}
\end{definition}




Now, let us derive essential lemmas that will be used in the following analysis. Note that inequalities used on matrix represent generalised inequality, e.g., $A \ge B$ implies that $A-B$ is positive semi-definite.





\begin{lemma}
\label{lm:spectral_radius}
For a positive semi-definite matrix $A\in\mathbb{R}^{m\times m}$, $A\leq I$ if and only if $\rho(A)\leq 1$, where the spectral radius $\rho(\cdot)$ is defined as the largest absolute value of the eigenvalues of a matrix.
\end{lemma}
\begin{pf}
For a positive semi-definite matrix $A\in\mathbb{R}^{m\times m}$, the maximum eigenvalue is the same as the spectral radius as:
\begin{equation}
\rho(A)= \max_{x\neq 0} \cfrac{x^TAx}{x^Tx}.
\end{equation}

For any non-zero vector $x\in\mathbb{R}^m$, the following inequality holds:
\begin{equation}
x^T(A-\rho(A)I)x = x^TAx - \rho(A)x^Tx \leq 0.
\end{equation}

Thus, the matrix $(A-\rho(A)I)$ being positive semi-definite, the following equation is satisfied:
\begin{equation}
\label{eq1.5}
A\leq \rho(A)I.
\end{equation}

The condition $A\leq I$ is satisfied if and only if $\rho(A) \leq 1$ in \eqn{eq1.5} \cite{Horn1985}.
\end{pf}

\begin{lemma}
\label{lm:bound_F}
For the DF$^2$ algorithm provided in \eqn{eq1.4}, the forgetting matrix $F(t)$ is bounded from above and below as:
\begin{equation}
\mu I \leq F(t) \leq I, \quad\forall t \ge 0.
\end{equation}
\end{lemma}
\begin{pf}

For DF$^2$, let us define a matrix $M(t) \triangleq I- F(t)$, i.e.:
\begin{equation}
M(t)=(1-\mu)\cfrac{P^{-1}(t-1)\phi(t)\phi^T(t)}{\phi^T(t)P^{-1}(t-1)\phi(t)}.
\end{equation}

Information matrix $R(t)$ is clearly positive semi-definite from \eqn{eq1.2} and thus $P(t)$ is also positive semi-definite for all $t$.  This implies that $M(t)$ is positive semi-definite and its eigenvalues are non-negative. The trace of $M(t)$ is obtained as:
\begin{equation}
tr(M(t)) = 1-\mu
\end{equation}

Since all eigenvalues of $M(t)$ are non-negative, the spectral radius of $M(t)$ is upper-bounded by $1-\mu$. By Lemma \ref{lm:spectral_radius}, DF$^2$ holds the following inequality:
\begin{equation}
0\leq M(t)\leq (1-\mu) I.
\end{equation}

Since this holds for all $t>0$ and $F(t)=I-M(t)$, the forgetting matrix $F(t)$ is bounded as:
\begin{equation}
\mu I \leq F(t) \leq I, \quad\forall t.
\end{equation}
\end{pf}

\begin{lemma}
\label{lm:bound}
For the forgetting algorithm of DF$^2$ in the form of \eqn{eq1.2}, the following inequality holds: 
\begin{equation}
\phi^T(t)P(t)\phi(t)<1, \quad \forall t. 
\end{equation}
\end{lemma}
\begin{pf}
It is clear from Lemma \ref{lm:bound_F} that $F(t)$ is invertible. Applying the matrix inversion lemma to \eqn{eq1.2}, the covariance matrix $P(t)$ is obtained as:
\begin{equation}
P(t)=\bar{P}(t)-\frac{\bar{P}(t)\phi(t)\phi^T(t)\bar{P}(t)}{1+\phi^T(t)\bar{P}(t)\phi(t)},
\end{equation}
where $\bar{P}(t)\triangleq P(t-1)F^{-1}(t)$. Rearranging the equation yields:
\begin{equation}
\phi^T(t)P(t)\phi(t) =\frac{\phi^T(t)\bar{P}(t)\phi(t)}{1+\phi^T(t)\bar{P}(t)\phi(t)}. 
\end{equation}

As $P(t)$ is positive semi-definite, $\bar{P}(t)$ is also positive semi-definite for all $t$. Hence, $\phi^T(t)P(t)\phi(t)<1$ is obtained for all $t>0$.
\end{pf}

\section{Analysis of Existing Forgetting Algorithms}
\label{sec:existing}
\subsection{Information Matrix Boundedness}

One of the key properties of the RLS algorithm is boundedness of the information matrix. Unboundedness from below increases the algorithm's sensitivity to noise, and unboundedness from above deteriorates the adaptation capability for time-varying parameters.

It is known that the EF algorithm cannot guarantee positive definiteness of the information matrix if the persistent excitation condition is not met \cite{Johnstone1982}. This implies that the EF algorithm is sensitive to noise, which results in the windup issue. The following lemma and proposition briefly review the boundedness of the information matrix in the EF algorithm.

\begin{lemma}[Boundedness of $R(t)$ in EF]
For EF, if the observed data is bounded, i.e., $\phi^T(t)\phi(t)\leq c$ for all $t$, the information matrix is bounded as
\begin{equation}
0\leq R(t) \leq \mu^{t} R(0) +\frac{1-\mu^t}{1-\mu}cI.
\end{equation}
\label{lm:bound_EF}
\end{lemma}

\begin{pf}
From \eqns{eq1.2} and \eqnref{eq1.4}, an explicit form of the information matrix in EF can be obtained  as:
\begin{equation}
R(t)=\mu^{t}R(0)+\sum_{i=0}^{t-1}\mu^i\phi(t-i)\phi^T(t-i).
\end{equation}

If the observed data is not persistently exciting, it is trivial that  $\lim\limits_{t\rightarrow \infty} R(t)=0$. Therefore, the lower bound of the information matrix is 0. From  $\phi^T(t)\phi(t)\leq c$ for all $t$, the following upper bound can be obtained:
\begin{equation}
\begin{aligned}
R(t) &\leq \mu^{t}R(0)+\sum_{i=0}^{t-1}\mu^i cI\\
&=\mu^{t} R(0) +\frac{1-\mu^t}{1-\mu}cI.
\end{aligned}
\end{equation}
\end{pf}

\begin{proposition}
\label{prop:bound_EF}
If the observed data is persistently exciting, the information matrix in EF is bounded from above and below as:
\begin{equation}
    aI\leq R(t) \leq \mu^{t} R(0) +\frac{1-\mu^t}{1-\mu}cI.
\end{equation}
where $a>0$ is a constant.
\end{proposition}

\begin{pf}
The proof is given as Lemma 1 in \cite{Johnstone1982}.

\end{pf}

Bittanti {\it et al.} \cite{Bittanti1990convergence} analysed the boundedness of DF$^1$: the lower boundedness of the information matrix in DF$^1$ is proven only with the PE condition satisfied. However, the upper-boundedness of the information matrix in DF$^1$ is not guaranteed even under the PE condition. 

\begin{lemma}[Boundedness of $R(t)$ in DF$^1$]
Suppose the observed data is persistently exciting and bounded, i.e. $\phi^T(t)\phi(t)\le c$. Then, the information matrix in DF$^1$ is bounded from below as:
\begin{equation}
\label{eq3.0}
R(t)\geq aI , 
\end{equation}
where $a>0$ is a constant.
\label{lm:bound_DF1}
\end{lemma}

\begin{pf}
The lower bound of the information matrix in DF$^1$ has been proven from Theorem 1 in \cite{Bittanti1990convergence}. 
\end{pf}

The boundedness of the information matrix in DF$^2$ has been proven in \cite{Cao2000}, but a brief proof on the upper boundedness is reviewed in the following lemma to support Remark \ref{remark:DF}.

\begin{lemma}[Boundedness of $R(t)$ in DF$^2$]
Suppose that the observed data is bounded, i.e., $\phi^T(t)\phi(t)\leq c$, $\forall t$. Then, the information matrix in DF$^2$ is bounded from above and below as:
\begin{equation}
\label{eq3.0.5}
aI \le R(t)\le bI , 
\end{equation}
where $a>0$ and $b>0$ are constant.
\label{lm:bound_DF2}
\end{lemma}

\begin{pf}
The lower bound of the information matrix in DF$^2$ has been proven in \cite{Cao2000}.

To obtain the upper bound of the information matrix, trace of $R(t)$ is computed as:
\begin{equation}
\begin{aligned}
tr(R(t))&=tr(R(0))\\
&+\sum_{i=1}^t\cfrac{\phi^T(i)\left(|\phi(i)|^2 I-\mu R(i-1)\right)R(i-1)\phi(i)}{\phi^TR(i-1)\phi(i)}.
\end{aligned}
\label{eq2.2}
\end{equation}

Suppose that the information matrix is not bounded from above, i.e. $\lambda(R(t))\rightarrow \infty$. Then, the left-hand-side of \eqn{eq2.2} is positive and unbounded, but the right-hand-side of \eqn{eq2.2} is negative. This contradicts the assumption. Therefore, the information matrix of DF$^2$ is bounded from above.
\end{pf}

\begin{remark}
\label{remark:DF}
Lemma \ref{lm:bound_DF2} shows that the information matrix is upper-bounded in DF$^2$, but does not show any actual bound or influence of $\mu$ in the upper bound. Recalling that the upper bound of the information matrix is closely related to the adaptation performance, this implies that it is difficult to investigate or control the adaptation performance in both DF$^1$ and DF$^2$.
\end{remark}

\subsection{Stability Characteristics}
\label{sec:3.2}

Another key property of the RLS algorithm is stability of the estimation error. This subsection investigates the stability characteristics of the selected benchmark forgetting algorithms, with and without PE condition. Like in many previous studies \cite{Johnstone1982,Bittanti1990convergence,Campi1992}, this paper adapts the following Lyapunov candidate function for the stability analysis: 
\begin{equation}
\label{eq3.1}
V(\tilde{\theta},t)=\cfrac{1}{2}\tilde{\theta}^T(t)R(t)\tilde{\theta}(t).
\end{equation}
Note that as discussed in \cite{Johnstone1982}, the boundedness of $R(t)$, i.e. $P(t)$, is of critical importance in the stability analysis: if the matrix is not bounded, especially from below, stability analysis based on this Lyapunov function might not be concluded. For brevity, we use $V(t)$ to denote $V(\tilde{\theta},t)$ unless it is necessary to keep the original notation. 

\begin{theorem}[Stability in EF]
\label{thm3.1}
If the observed data is persistently exciting, the equilibrium point $\tilde{\theta}(t) \equiv 0$ in EF is exponentially stable.
\end{theorem}
\begin{pf}
Exponential stability under PE condition is proven as Theorem 1 in \cite{Johnstone1982}.
\end{pf}

\begin{remark}
\label{remark:EF}
Since the lower bound of $R(t)$ in EF cannot be determined without the PE condition satisfied. The positive definiteness of the Lyapunov function cannot be guaranteed. This implies that the stability of the equilibrium point $\tilde{\theta}(t) \equiv 0$ cannot be concluded. 
\end{remark}

Under the PE condition, stability in DF$^1$ has been proven in \cite{Bittanti1990convergence}, but the stability characteristics in DF$^2$ has not been analysed. This paper performs stability analysis of DF$^2$ for the first time. The following theorems briefly review the stability characteristics in DF$^1$ and analyse DF$^2$.

\begin{theorem}[Stability in DF$^1$]
\label{thm3.2}
If the observed data is persistently exciting, the equilibrium point $\tilde{\theta}(t) \equiv 0$ in DF$^1$ is stable.
\end{theorem}
\begin{pf}
Under the PE condition satisfied, stability in DF$^1$ is proven as Theorem 2 in \cite{Bittanti1990convergence}.
\end{pf}

\begin{theorem}[Stability in DF$^2$]
\label{thm3.3}
The equilibrium point $\tilde{\theta}(t) \equiv 0$ in DF$^2$ is uniformly stable, regardless of the PE condition.
\end{theorem}
\begin{pf}
Consider the Lyapunov candidate function given in \eqn{eq3.1}. Lemma \ref{lm:bound_DF2} shows that the information matrix in DF$^2$ is bounded from below and above regardless of PE condition. From the lower bound given in Lemma \ref{lm:bound_DF2}, it is clear that $V(0,t) = 0$ and $V(\tilde{\theta},0)>0$ for all $\tilde{\theta} \neq 0$. Moreover, from \eqn{eq3.0.5}, we have
\begin{equation}
    V(\tilde{\theta},t) \ge a || \tilde{\theta} ||
\end{equation}
Therefore, the Lyapunov function is positive definite.

The upper bound given in \eqn{eq3.0.5} yields:
\begin{equation}
   V(\tilde{\theta},t) \le b || \tilde{\theta} || 
\end{equation}
Hence, the Lyapunov function is decrescent. 

From \eqns{eq1.1} and \eqnref{eq1.2}, the Lyapunov function can be computed as:
\begin{equation}
\resizebox{.88\hsize}{!}{$
\begin{aligned}
&V(t)=\cfrac{1}{2}\tilde{\theta}^T(t-1)F(t)R(t-1)\tilde{\theta}(t-1)\\
&-\cfrac{1}{2}\left[1-\phi^T(t)P(t)\phi(t) \right] \tilde{\theta}^T(t-1)\phi(t)\phi^T(t)\tilde{\theta}(t-1) .
\end{aligned}$}
\end{equation}

From Lemmas \ref{lm:bound_F} and \ref{lm:bound}, $1-\phi^T(t)P(t)\phi(t)>0$ and $F(t)\leq I$ for all $t$. Therefore, we have:
\begin{equation}
\begin{aligned}
V(t)&\leq \cfrac{1}{2}\tilde{\theta}^T(t-1)R(t-1)\tilde{\theta}(t-1) \\
&\leq V(t-1).
\end{aligned}
\end{equation}

Since $V(t)\leq V(t-1)$ for all $t$ and $V(t)$ is decrescent positive definite, the equilibrium point $\tilde{\theta}(t) \equiv 0$ is uniformly stable.
\end{pf}

\begin{remark}
Given that the observed data is persistently exciting, EF guarantees exponential stability whereas DF$^1$ and DF$^2$ guarantee stability and uniform stability, respectively. It can be inferred that the convergence speed of the EF algorithm might be faster than the DF algorithms with the PE condition satisfied.
\end{remark}

\section{New EF Algorithm}
\label{sec:newEF}
This section develops a new EF algorithm. The main objective of the proposed algorithm is to guarantee a lower bound of the information matrix to mitigate the estimator windup issue and retain fast convergence speed of the EF algorithm. In the proposed EF algorithm, the adaptation gain and the forgetting matrix are defined as:
\begin{align}
K(t) &= P(t) \\
F(t) &=  \mu I + \delta P(t-1),\label{eq4.1}
\end{align}
where $\delta > 0$ is a design parameter. Note that we introduce an additional term $\delta P(t-1)$ to the forgetting matrix $F(t)$ of the original EF. This is to achieve the boundedness of the information matrix, while maintaining the convergence speed of the EF algorithm. 

\subsection{Information Matrix Boundedness}
The following two theorems show that the information matrix in the new EF algorithm is bounded from above and below.

\begin{theorem}[Lower boundedness of $R(t)$]
\label{thm4.1}
Consider the update \eqns{eq1.2} and \eqnref{eq4.1}. Suppose the observed data is persistently exciting in the order of $s$, and there exists a positive $\delta$ that satisfies:
    \begin{equation}
    \label{eq4.3}
    \begin{aligned}
    \delta \leq (1-\mu)\Big(\lambda_{min}(R(t))-&\cfrac{\sum_{i=1}^{t}\mu^i \phi^T(t)\phi(t) }{1-\mu^t}\Big), \\
    &\text{ for } 0\leq t <s.
    \end{aligned}
    \end{equation}
Then,
    \begin{equation}
    R(t) \geq \cfrac{\delta}{1-\mu}I+\cfrac{\sum_{i=0}^{s-1}\mu^i \phi(t-i)\phi^T(t-i) }{1-\mu^s}, \quad \forall t \; .
    \end{equation}
\end{theorem}
\begin{pf}
For $0\leq t<s$, the condition on $\delta$ yields:
\begin{equation}
\lambda_{min}(R(t))\geq \frac{\delta}{1-\mu}+ \frac{\sum_{i=1}^{t}\mu^i \phi(t)\phi^T(t)}{1-\mu^t}.
\end{equation}

Thus, it is clear that:
\begin{equation}
\label{eq4.4}
R(t) \geq \cfrac{\delta}{1-\mu}I+\cfrac{\sum_{i=1}^{t}\mu^i \phi(t)\phi^T(t) }{1-\mu^t}\text{ for } 0\leq t<s.
\end{equation}

Assuming that the following condition holds for some $t\geq s$:
\begin{equation}
\label{eq4.5}
R(t-s) \geq \cfrac{\delta}{1-\mu}I+\cfrac{\sum_{i=0}^{s-1}\mu^i \phi(t-i)\phi^T(t-i)}{1-\mu^s},
\end{equation}
we have:
    \begin{equation}
    \resizebox{.88\hsize}{!}{$
    \begin{aligned}
    R (t) &= \mu^sR(t-s)+\sum_{i=0}^{s-1}\mu^i \left[ \phi(t-i)\phi^T(t-i)+\delta I \right] \\
    & \geq \cfrac{\delta}{1-\mu}I+\cfrac{\sum_{i=0}^{s-1}\mu^i \phi(t-i)\phi^T(t-i)}{1-\mu^s} \; .
    \end{aligned}   
    $}
    \end{equation}

Now, we can complete the proof using the concept of induction. \eqn{eq4.4} indicates that $R(t)$ is lower-bounded for $t \in [0, s)$.  If $R(t-s)$ is lower-bounded for all $t\geq s$, \eqn{eq4.5} shows that $R(t)$ is also lower-bounded for all $t\geq s$. Hence, following the concept of induction, we obtain:
    \begin{equation}
    R(t) \geq \cfrac{\delta}{1-\mu}I+\cfrac{\sum_{i=0}^{s-1}\mu^i \phi(t-i)\phi^T(t-i) }{1-\mu^s}, \quad \forall t.
    \end{equation}
\end{pf}

\begin{corollary}
\label{cor4.1}
Suppose the PE condition is not hold, and there exists a positive $\delta$ that satisfies:
\begin{equation}
\label{eq4.6}
\delta \leq(1-\mu)\lambda_{min}(R(0)).
\end{equation}

Then, 
\begin{equation}
R(t) \geq \cfrac{\delta}{1-\mu}I , \quad\forall t \; .
\end{equation}
\end{corollary}
\begin{pf}
Similar to Theorem \ref{thm4.1}, the lower bound of the information matrix can be proven by induction even for the case where the PE condition is not satisfied. From the condition on $\delta$ in \eqn{eq4.6} yields:
\begin{equation}
R(0)\geq \cfrac{\delta}{1-\mu}I.
\end{equation}

From \eqns{eq1.2} and \eqnref{eq4.1}, the update of the information matrix can be written as:
\begin{equation}
\label{eq4.1.5}
R(t) = \mu R(t-1)+\phi(t)\phi^T(t) + \delta I.
\end{equation}

Assuming that the following condition holds for some $t> 0$:
\begin{equation}
R(t-1)\geq \cfrac{\delta}{1-\mu}I,
\end{equation}
we have:
\begin{equation}
\begin{aligned}
R(t)&=\mu R(t-1)+\phi(t)\phi^T(t) + \delta I \\
&\ge \cfrac{\delta\mu}{1-\mu}I + \delta I.
\end{aligned}
\end{equation}

Hence, the lower boundedness of $R(t-1)$ for $t>0$ leads to the lower boundedness of $R(t)$. From induction, the following lower bound is satisfied: 
\begin{equation}
R(t) \geq \cfrac{\delta}{1-\mu}I , \quad\forall t \; .
\end{equation}
\end{pf}

\begin{remark}
\label{rem4.1}

It is known that the EF algorithm could suffer from the estimator windup problem as the covariance matrix cannot be uniformly bounded from above without PE, i.e. the information matrix cannot be uniformly bounded from below. In the proposed EF algorithm, Corollary \ref{cor4.1} shows that the covariance matrix is uniformly bounded from above even without PE. The lower bound of the information matrix is determined by tuning $\delta$ and $\mu$. This implies that the lower bound, and thus the sensitivity bound, can be controlled by $\delta$ and $\mu$ in the proposed algorithm. \eqn{eq4.6} clearly shows that $R(0)$ should be properly chosen to determine $\delta$ and consequently the lower bound. 
\end{remark}

\begin{theorem}[Upper boundedness of $R(t)$]
\label{thm4.2}
Suppose that $\phi^T(t)\phi(t) \le c$ for all $t$. Then, the following condition holds:
\begin{equation}
R(t) \le \mu^t R(0) +\cfrac{1-\mu^t}{1-\mu}(c+\delta)I .
\end{equation}
\end{theorem}

\begin{pf}
From \eqns{eq1.2} and \eqnref{eq4.1}, $R(t)$ can be expressed as:
    \begin{equation}
    R(t)=\mu^t R(0)+\sum_{i=0}^{t-1}\mu^i (\phi(t-i)\phi^T(t-i) + \delta I).
    \end{equation}

As $\phi^T(t)\phi(t) \le c$ for all $t$, we have: 
    \begin{equation}
    R(t) \leq \mu^t R(0) +\cfrac{1-\mu^t}{1-\mu}(c+\delta)I .
    \end{equation}
\end{pf}

\begin{remark}
\label{rem4.2}
Theorem \ref{thm4.2} shows the uniform positiveness of the covariance matrix in the proposed approach. Roughly speaking, this means that the information contents of all the parameters do not tend to infinity. Therefore, Theorem \ref{thm4.2} implies that the proposed algorithm can retain the responsiveness to parameter variations in a certain degree. Recall that Remark \ref{remark:DF} states that some DF algorithms also guarantee the upper boundedness of $R(t)$, but it is difficult to investigate the adaptation performance since the actual bound is not obtained. Unlike in the DF algorithms, $\delta$ and $\mu$ can be directly determined to design the upper bound of $R(t)$ and thus the responsiveness to the parameter change.
\end{remark}

\subsection{Stability Characteristics}
Now, let us examine stability of the proposed algorithm.

\begin{theorem}
\label{thm4.3}
Consider the proposed EF algorithm with $\delta$ satisfying \eqn{eq4.3}. If the observed data is persistently exciting, the  equilibrium point $\tilde{\theta}(t) \equiv 0$ in the proposed EF is exponentially stable. If the PE condition is not met, the equilibrium point is uniformly stable.
\end{theorem}
\begin{pf}
Consider the Lyapunov candidate function given in \eqn{eq3.1}, i.e.:
\begin{equation}
\begin{aligned}
V(t)&=\cfrac{1}{2}\tilde{\theta}^T(t)R(t)\tilde{\theta}(t)\\
&\leq \cfrac{1}{2}\tilde{\theta}^T(t-1)F(t)R(t-1)\tilde{\theta}(t-1).
\end{aligned}
\end{equation}
The information matrix is bounded from above and below as proven in Corollary \ref{cor4.1} and Theorem \ref{thm4.2} regardless of the PE condition. As shown in Theorem \ref{thm3.3}, this indicates that the Lyapunov function is positive definite and decrescent.

If the observed data is persistently exciting, Theorem \ref{thm4.1} yields that:
\begin{equation}
R(t) > \cfrac{\delta}{1-\mu}I  \; \Rightarrow \; \delta I < (1 - \mu) R(t) \quad \forall t.
\end{equation}
This yields:
\begin{equation}
F(t)R(t-1)= \mu R(t-1) + \delta I <R(t-1) \quad \forall t.
\end{equation}
Hence, $V(t)<V(t-1)$ holds for all $t$, and thus the equilibrium point $\tilde{\theta}(t) \equiv 0$ is exponentially stable. 

If the PE condition is not met, $F(t)R(t-1)\leq R(t-1)$ for all $t$ holds from Corollary \ref{cor4.1}. Since $V(t)$ is a decrescent positive definite function and $V(t)\leq V(t-1)$, the proposed EF algorithm guarantees uniform stability of the equilibrium point $\tilde{\theta}(t) \equiv 0$.
\end{pf}

\begin{remark}
\label{rem4.3}
As shown in Theorem \ref{thm3.1} and discussed in Remark \ref{remark:EF}, the EF algorithm cannot guarantee any type of stability without the PE condition satisfied. Unlike the EF algorithm, the proposed EF algorithm can guarantee uniform stability even without the PE condition.  
\end{remark}

\begin{remark}
\label{rem4.4}
As discussed in Introduction, fast convergence is desirable as it implies adaptation capability for time-varying parameters: this could become of paramount importance in many applications. Theorems \ref{thm3.2}, \ref{thm3.3}, and \ref{thm4.3} shows that the proposed EF algorithm guarantees either uniform or exponential stability depending on the PE condition, whereas the DF algorithms guarantee stability or uniform stability. We can infer that the convergence speed of the proposed EF algorithm could be faster than that of the DF algorithms compared. 
\end{remark}

The summary of the analysis on information matrix boundedness and stability characteristics is shown in \tab{tab:summary}. For boundedness of $R(t)$, the acronyms LB, UB, and LUB stand for lower-bounded, upper-bounded, and lower and upper-bounded, respectively. For stability characteristics, {\it exponential} and {\it uniform} stand for exponential stability and uniform stability, respectively. The acronyms Prop, Thm, and Cor represent Proposition, Theorem, and Corollary.

\begin{table}[t]
\renewcommand*{\arraystretch}{1.3}
    \centering
    \caption{Summary of Analysis Results}
    \resizebox{\hsize}{!}{
    \begin{tabular}{lllll}
        \hline
         & \multicolumn{2}{l}{Boundedness of $R(t)$} & \multicolumn{2}{l}{Stability Characteristics}  \\
         & Without PE & With PE & Without PE & With PE\\
         \hline
         EF & UB & LUB & - & Exponential \\
         & (Lemma \ref{lm:bound_EF}) & (Prop \ref{prop:bound_EF}) &  & (Thm \ref{thm3.1})\\
         DF$^1$ & - & LB & - & Stability \\
         &  & (Lemma \ref{lm:bound_DF1}) &  & (Thm \ref{thm3.2})\\
         DF$^2$ & LUB & LUB & Uniform & Uniform \\
         & (Lemma \ref{lm:bound_DF2})& (Lemma \ref{lm:bound_DF2}) & (Thm \ref{thm3.3}) & (Thm \ref{thm3.3}) \\
         Proposed & LUB & LUB & Uniform & Exponential\\
         EF & (Cor \ref{cor4.1} \& Thm \ref{thm4.2})  & (Thm \ref{thm4.1} \& \ref{thm4.2})  & (Thm \ref{thm4.3}) & (Thm \ref{thm4.3})\\
         \hline
    \end{tabular}}
    \label{tab:summary}
\end{table}


\section{Numerical Simulation}
\label{sec:NS}
This section performs numerical simulations to validate the theoretical analysis results. For rigorous validation, the performance of the proposed algorithm is compared with the benchmark EF, DF$^1$ and DF$^2$ algorithms. 

\subsection{Simulation Setup}
The proposed EF algorithm is numerically investigated with wing rock phenomenon, which is continuous lateral oscillations commonly present in highly swept-back or delta wing aircrafts. The wing rock roll dynamics is widely utilised as an example to estimate the parameters in many existing literature, for its uncertainty in the nonlinear dynamics of wing rock motion and aeroelastic systems. Various mathematical models have been developed to describe the wing rock, among which the model developed by Singh {\it et al.} \cite{Singh1995} is considered for its accuracy and simplicity. The dynamics is described as:
%
%
\begin{equation}
\dot{x}_1 = x_2, \quad 
\dot{x}_2 =\Delta(x)+L_{\delta_a}\delta_a
\end{equation}
where $x_1$ and $x_2$ are the roll angle and its rate, $\delta_a$ the aileron deflection, and $L_{\delta_a}(=1)$ the effectiveness. 

Defining the state vector as $x=[x_1\quad x_2]^T$, the uncertainty $\Delta(x)$ is structured as:
\begin{equation}
\Delta(x) = \phi^T(x)\theta(t),
\end{equation}
where the observed data is set as:
\begin{equation}
\phi(x)=[1, x_1, x_2, |x_1|x_2, |x_2|x_1,x_1^3]^T.
\end{equation}

This paper considers two cases to investigate three main aspects, i.e., adaptation capability, estimation windup issue and stability characteristics. The parameters in the two cases are given by:
\begin{equation}
\resizebox{.88\hsize}{!}{$
\begin{aligned}
    \text{C1: }& \theta(t)=\begin{cases} 
[.8, .2314, .6918, -.6245, .0095, .0214]^T, \; t<50 \\
[.88, .2198, .6295, 1.1856, .0114, .0208]^T, \; t\geq 50
\end{cases} \\
\text{C2: } & \theta(t)=
[.8, .2314, .6918, -.6245, .0095, .0214]^T.
\end{aligned}
$}
\end{equation}
In case 1, we inject an aerodynamic parameter change at $t=50$ for investigating adaptation capability. Note that the aerodynamic parameters $\theta(t)$ are given in Singh {\it et al.} \cite{Singh1995}. In order to check estimation windup issue, the noise is also injected at $t\geq 60$ sec in the first case. The noise is generated by Gaussian distribution with variance 0.1 and zero mean. In case 2, neither parameter change nor noise is injected in the simulation to solely investigate on the stability characteristics.

The simulation is conducted with the time step 0.01 sec, with the initial covariance matrix is set as $I_{6\times 6}$. For EF algorithms, $\mu$ and $\delta$ are set as 0.99 and 0.01, respectively. For DF algorithms, $\mu$ is set as 0.95 to achieve comparable convergence speed. The aileron deflection $\delta_a$ is designed as \cite{Chowdhary2014}:
\begin{equation}
\delta_a(t) = K_p(r(t)-x_1(t))-K_dx_2(t).
\end{equation}
where the linear gains $K_p$ and $K_d$ are set 1.5 and 1.3, respectively, and $r(t)$ is the reference input.

\subsection{Simulation Results}

\begin{figure}[t!]
\centering
\includegraphics[width=\hsize]{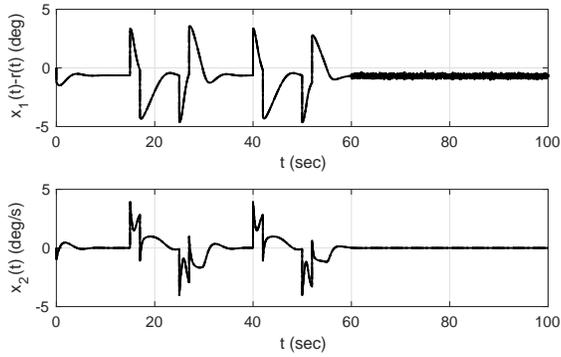}
\caption{State vector in case 1}
\label{fig:State}
\end{figure}
\begin{figure}[t!]
\centering
\begin{subfigure}[t]{\hsize}
    \includegraphics[width=\hsize]{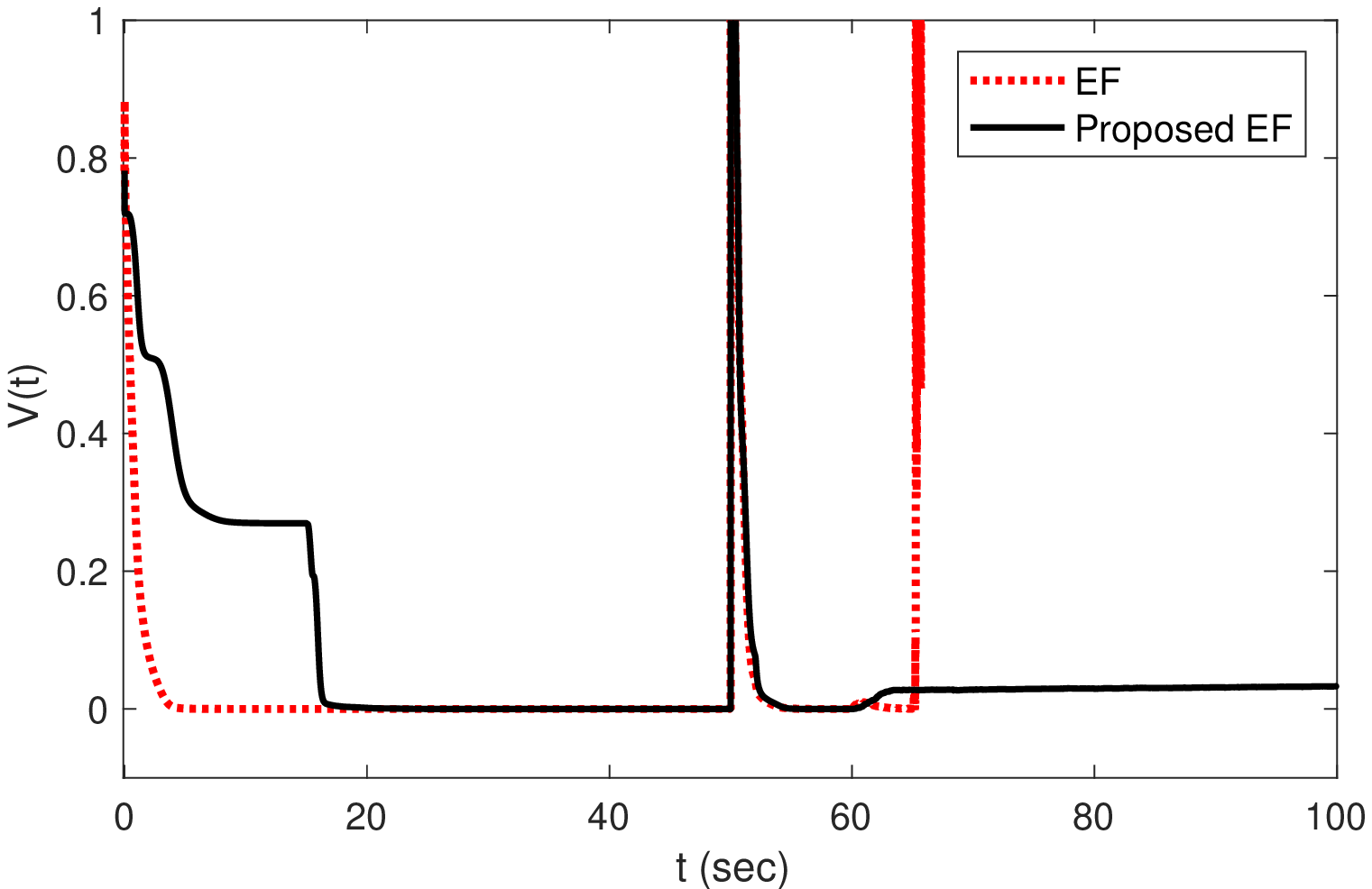}
    \caption{EF algorithms}
\end{subfigure}
\begin{subfigure}[t]{\hsize}
    \includegraphics[width=\hsize]{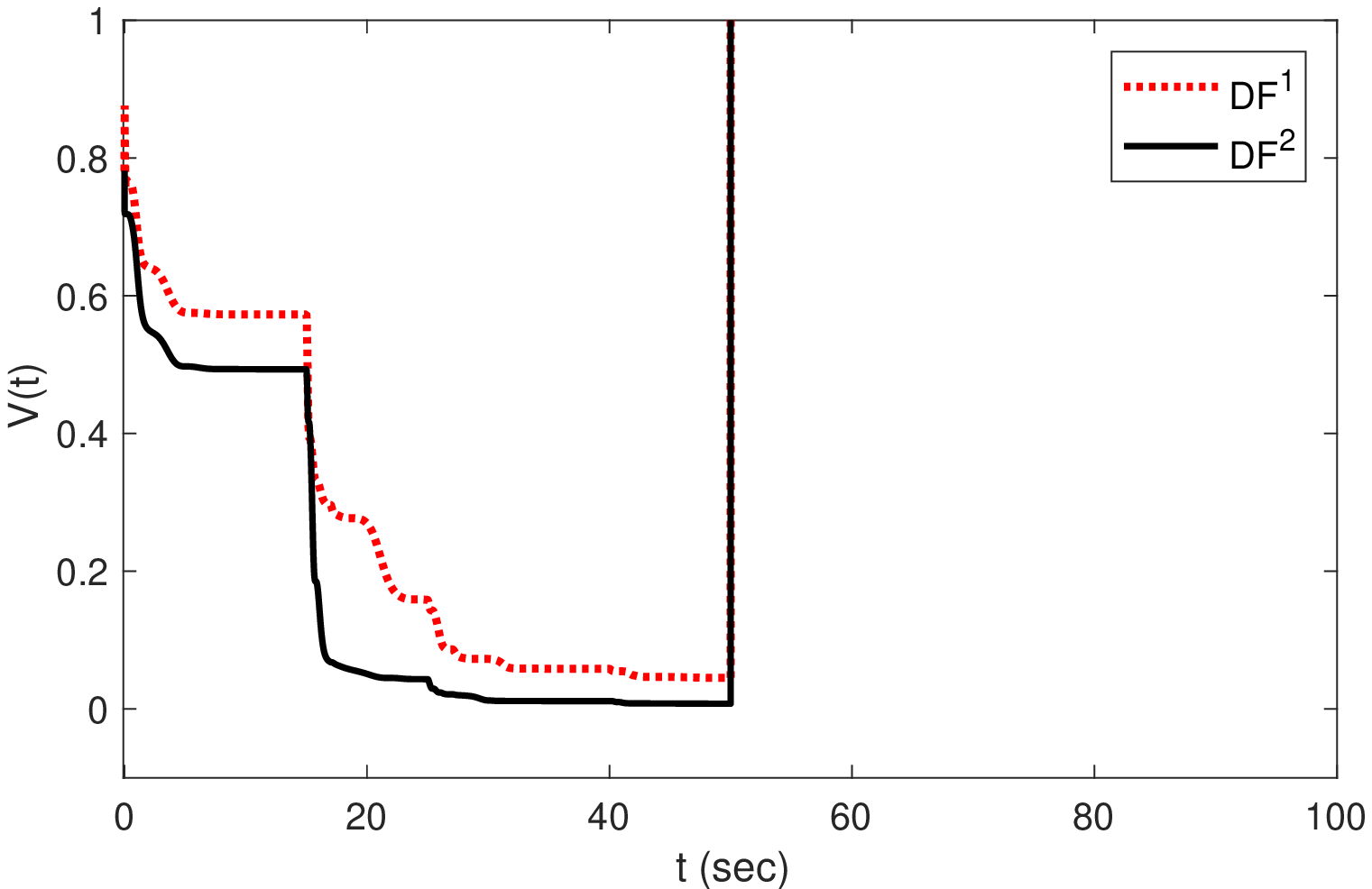}
    \caption{DF algorithms}
\end{subfigure}
\caption{Lypunov function in case 1}
\label{fig:Lyapunov}
\end{figure}
\begin{figure}[t!]
\centering
\begin{subfigure}[t]{\hsize}
    \includegraphics[width=\hsize]{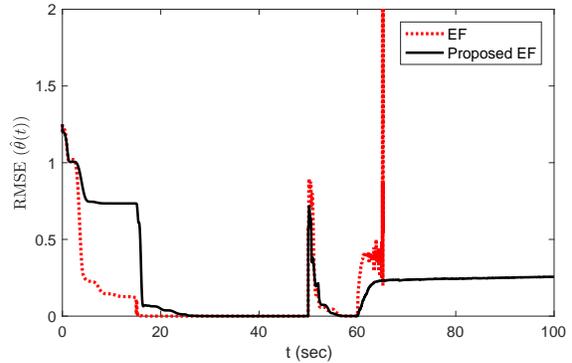}
    \caption{EF algorithms}
\end{subfigure}
\begin{subfigure}[t]{\hsize}
    \includegraphics[width=\hsize]{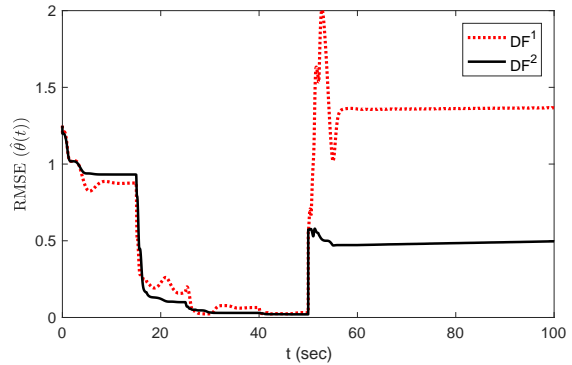}
    \caption{DF algorithms}
\end{subfigure}
\caption{Estimation error in case 1}
\label{fig:Error}
\end{figure}

\fig{fig:State} shows time histories of state values of simulation case 1. As depicted in \fig{fig:State}, there exists no excitation except the noise after around 60 sec. The profiles of the Lyapunov function and root-mean-square error of $\hat{\theta}(t)$ with different forgetting algorithms are shown in \figs{fig:Lyapunov} and \ref{fig:Error}. Note that the as expected, profiles of the Lyapunov function and estimation error show same tendency.

\figs{fig:Lyapunov} and \ref{fig:Error} show that the proposed EF algorithm is the only algorithm whose Lyapunov function and estimation error are both stable. With the parameter variation at $t= 50$ sec, the proposed EF algorithm and conventional EF algorithm converge much faster than the DF algorithms. These results confirm that the proposed EF algorithm could provide stronger adaptation capability to the time-varying parameters, compared with the DF algorithms. This validates Remark \ref{rem4.2}. Under the existence of noise over $t \ge 60$ sec, the Lyapunov function and estimation error of the proposed EF algorithm stay bounded, whereas the Lyapunov function and estimation error diverge in the conventional EF algorithm. This demonstrates that the proposed algorithm can alleviate the estimation windup issue unlike the conventional EF algorithm, complying with Remark \ref{rem4.1}. Note that the Lyapunov function in the DF algorithms also stays bounded, but above the value of $1$, and hence their estimation error is bounded even after 50 sec as shown in \fig{fig:Error}. The simulation results show that the tracking capability of the proposed EF algorithm is better than that of DF algorithms: unlike DF algorithms, the estimation error in the proposed algorithm is converged after the parameter change at 50 sec, before the noise is injected at 60 sec. 

\begin{figure}[t!]
\centering
\includegraphics[width=\hsize]{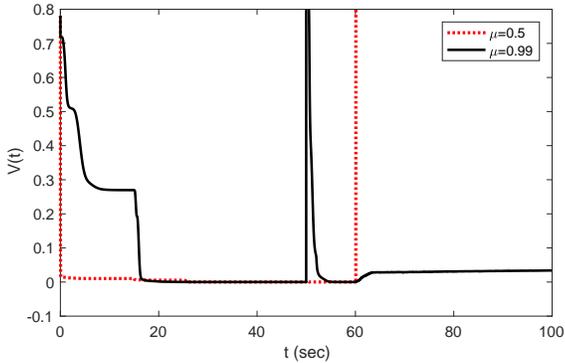}
\caption{Lypunov function with different $\mu$ in case 1}
\label{fig:Effect1}
\end{figure}

\begin{figure}[t!]
\centering
\includegraphics[width=\hsize]{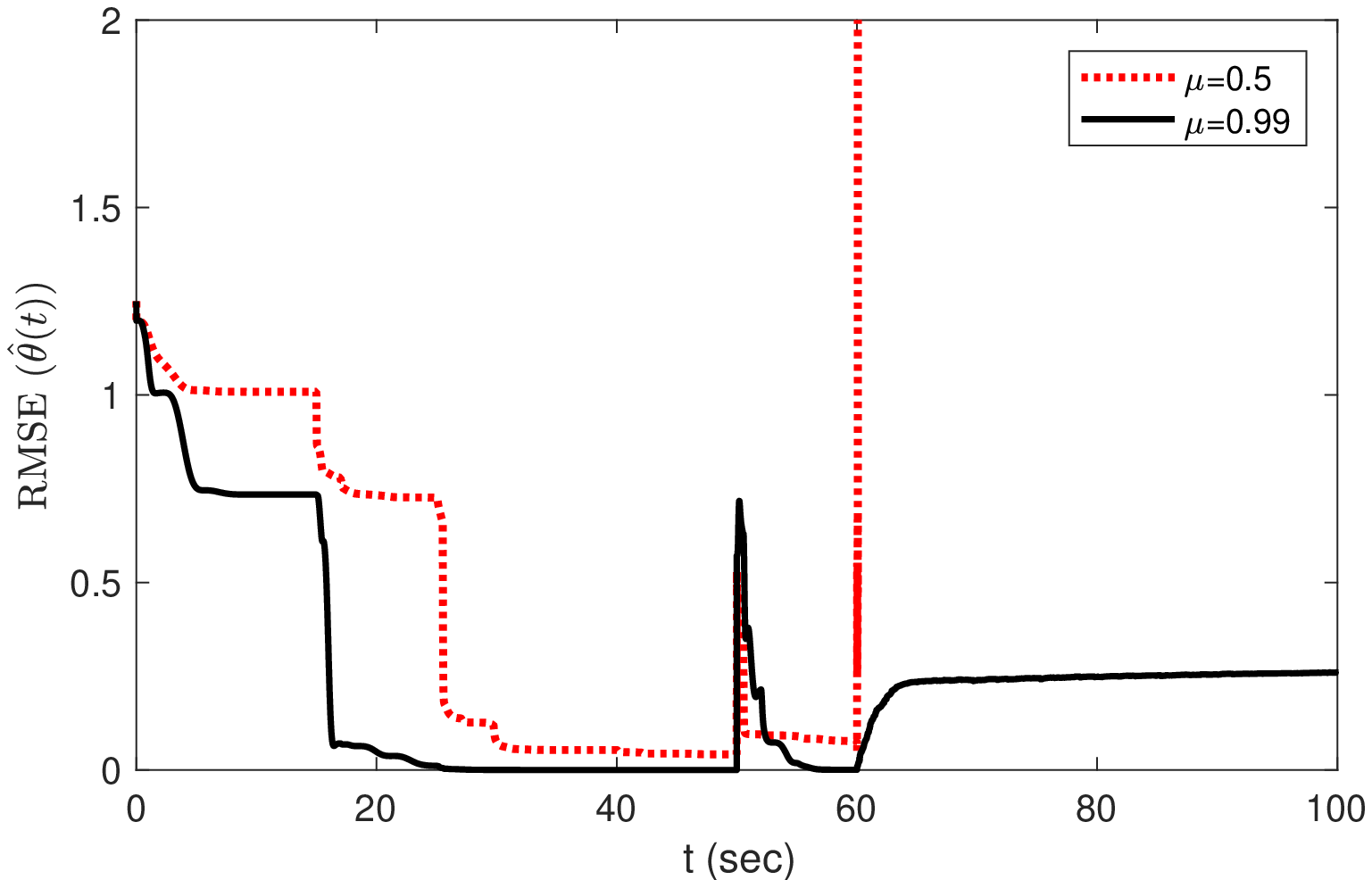}
\caption{Estimation error with different $\mu$ in case 1}
\label{fig:Effect2}
\end{figure}

\fig{fig:Effect1} and \ref{fig:Effect2} show the Lyapunov function and estimation error of the proposed EF algorithm, with different values of $\mu$, 0.5 and 0.99. Since increase in $\mu$ leads to higher lower bound of the information matrix as shown in Corollary \ref{cor4.1}, the proposed EF algorithm with higher $\mu$ is less sensitive to the noise, resolving the estimation windup issue. On the other hand, the adaptation capability to the time-varying parameters is better with lower $\mu$, as the upper bound of the information matrix is lower from Theorem \ref{thm4.2}. The results shown in \fig{fig:Effect1} and \ref{fig:Effect2} are complied with these analysis results. This implies that we can design $\mu$ and $\delta$ to control the sensitivity bound and responsiveness to the parameter change, as discussed in Remark \ref{rem4.1} and \ref{rem4.2}.

\begin{figure}[t!]
\centering
\includegraphics[width=\hsize]{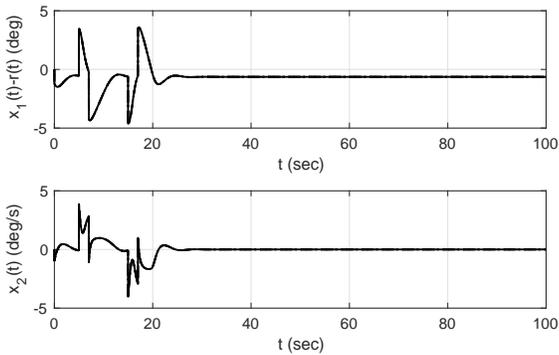}
\caption{State vector in case 2}
\label{fig:State2}
\end{figure}
\begin{figure}[t!]
\centering
\begin{subfigure}[t]{\hsize}
    \includegraphics[width=\hsize]{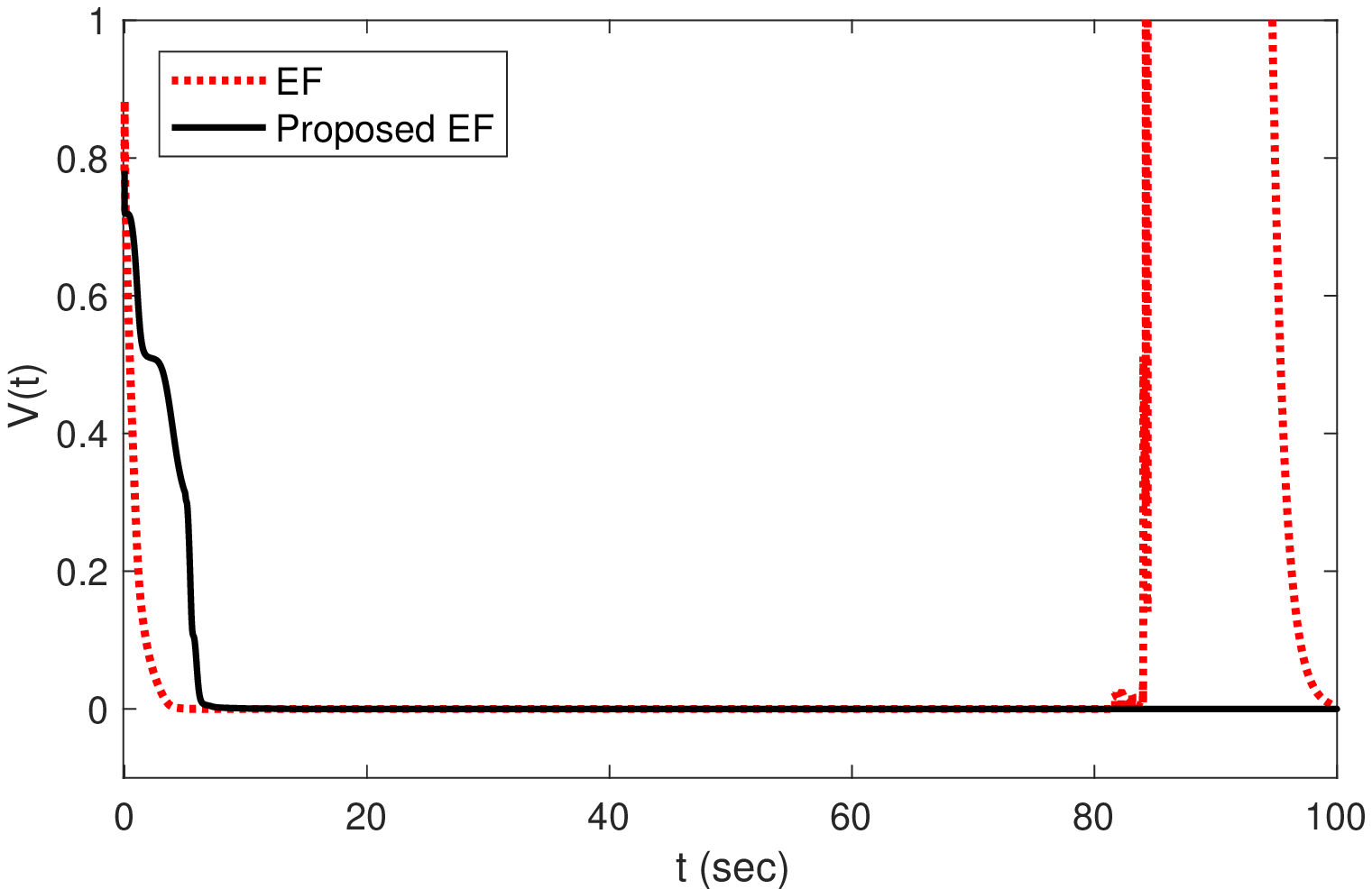}
    \caption{EF algorithms}
\end{subfigure}
\begin{subfigure}[t]{\hsize}
    \includegraphics[width=\hsize]{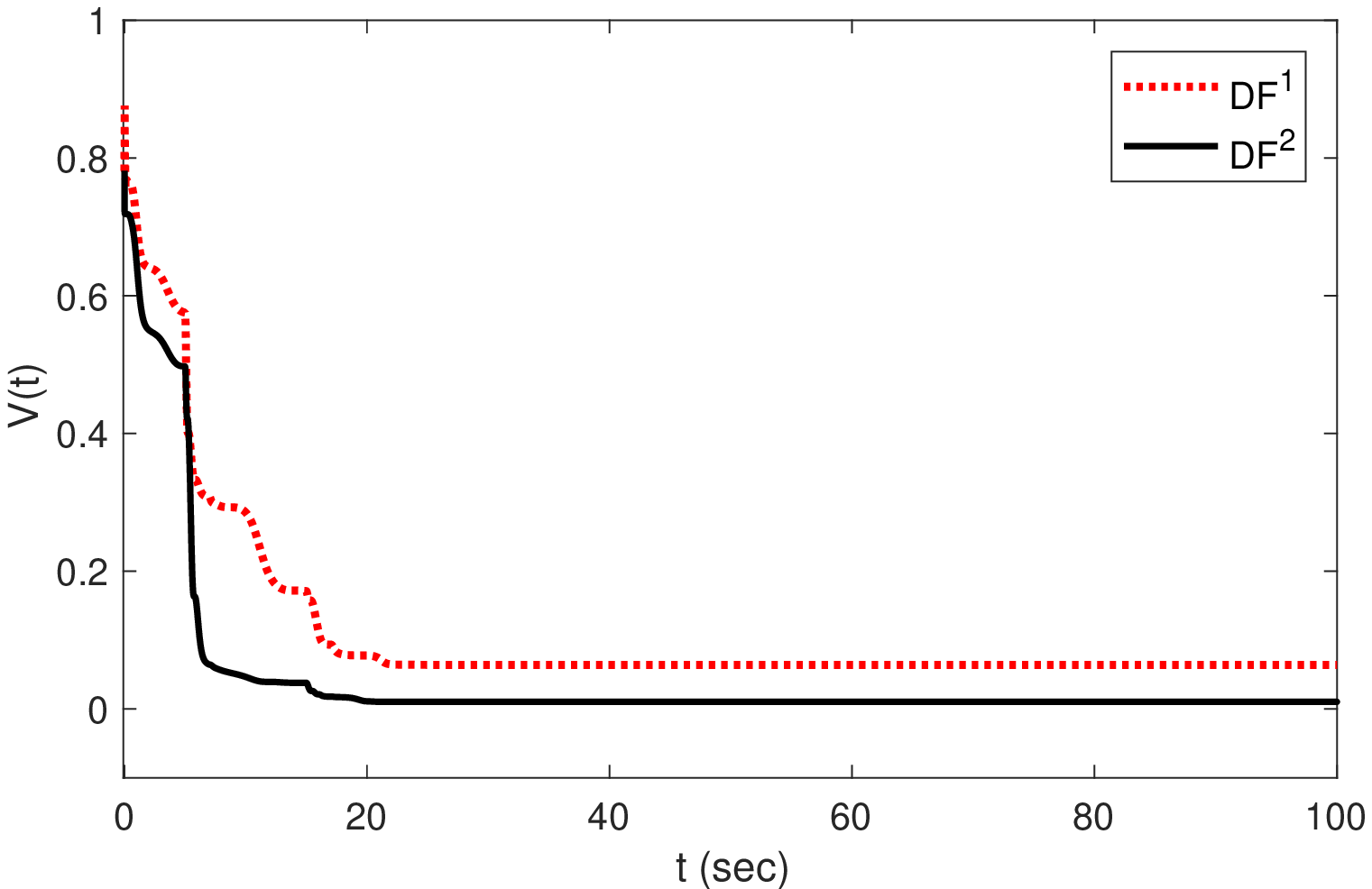}
    \caption{DF algorithms}
\end{subfigure}
\caption{Lypunov function in case 2}
\label{fig:Lyapunov2}
\end{figure}
\begin{figure}[t!]
\centering
\begin{subfigure}[t]{\hsize}
    \includegraphics[width=\hsize]{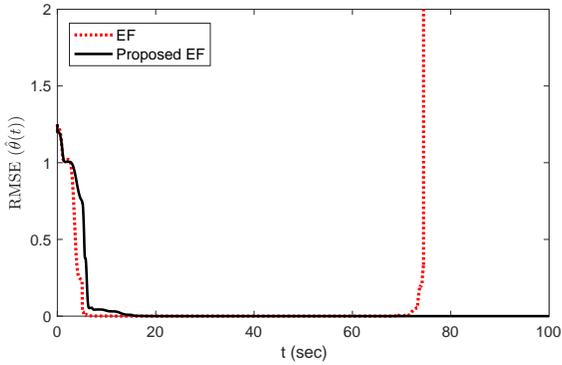}
    \caption{EF algorithms}
\end{subfigure}
\begin{subfigure}[t]{\hsize}
    \includegraphics[width=\hsize]{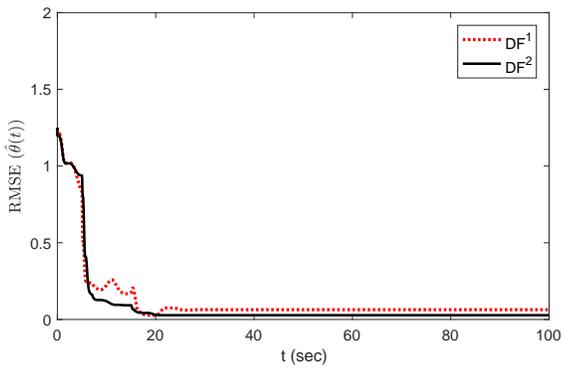}
    \caption{DF algorithms}
\end{subfigure}
\caption{Estimation error in case 2}
\label{fig:Error2}
\end{figure}

The state values of simulation case 2 are shown in \fig{fig:State2}. As shown in the figure, the signals are excited until around 30 sec, and there is no excitation afterwards. The profiles of the Lyapunov function and RMSE of $\hat{\theta}(t)$ for case 2 are shown in \figs{fig:Lyapunov2} and \ref{fig:Error2}, respectively. 

The results in \figs{fig:Lyapunov2} and \ref{fig:Error2} confirm that regardless of the PE condition, the Lyapunov function and thus estimation error of the proposed EF algorithm converge, whereas those of the conventional EF algorithm diverge without the PE condition. Although the estimation error remains divergent, \fig{fig:Lyapunov2} illustrates that the value of Lyapunov function converges to zero at around 100 sec in the EF algorithm. This is because the information matrix of the EF algorithm becomes zero without any excitation. The results comply with Remark \ref{rem4.3}. Also, under the PE condition, the Lyapunov function converges faster in EF algorithms than in DF algorithms, as discussed in Remark \ref{rem4.4}. Even if there is no excitation, the Lyapunov function of the proposed algorithm stays around $10^{-11}$, whereas it stays around 0.06 and 0.01 in DF$^1$ and DF$^2$.

\section{Conclusion}
\label{sec:conclusion}
This paper developed a new EF algorithm that can prevent typical issues with estimator windup and the stability without the PE condition, while retaining the adaptation capability of the EF algorithm. To identify potential issues with existing algorithms, this paper first extensively investigates stability properties and boundedness of the covariance matrix in various exponential and directional forgetting algorithms. The analysis of the proposed EF algorithm confirms that it guarantees exponential stability with PE and uniform stability without PE. Also, the analysis shows the information matrix of the proposed EF algorithm is bounded from above and below regardless of the PE condition. This implies that the new EF algorithm can prevent the estimator windup problem and at the same time maintaining the adaptation capability to time-varying parameters.


\bibliographystyle{plain}        
\bibliography{PE_EF}           



\end{document}